\documentclass{article}

\usepackage{arxiv}

\usepackage[utf8]{inputenc} 
\usepackage[T1]{fontenc}    
\usepackage{hyperref}       
\usepackage{url}            
\usepackage{booktabs}       
\usepackage{nicefrac}       
\usepackage{microtype}      
\usepackage{lipsum}		
\usepackage{graphicx}
\usepackage{doi}
\usepackage{subcaption}
\usepackage{tikz}
\usepackage{placeins}
\usepackage{multirow}
\usepackage{array,booktabs,xcolor}
\usepackage{caption}
\usepackage{url}
\usepackage{dsfont}
\usepackage{amsmath,amssymb, amsthm}
\usepackage{tablefootnote} 
\usepackage[ruled,vlined]{algorithm2e} 

\newtheorem{definition}{Definition}
\newtheorem{prop}{Proposition}
\UseRawInputEncoding 

\title{Cybersecurity threat detection based on a UEBA framework using Deep Autoencoders}

\author{ Jos\'e Fuentes \\
	Galician Research and Development Center \\
    in Advanced Telecommunications (Gradiant)\\       
    36214 Vigo, Spain\\
	\texttt{jfuentes@gradiant.org} \\
	\And
    \href{https://orcid.org/0000-0002-8041-6860}{\includegraphics[scale=0.06]{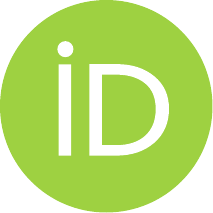}\hspace{1mm}Ines Ortega-Fern\'andez}\thanks{Corresponding author} \\
	Galician Research and Development Center \\
    in Advanced Telecommunications (Gradiant)\\ 
    36214 Vigo, Spain\\
	\texttt{iortega@gradiant.org} 
\And
\href{https://orcid.org/0000-0001-8085-2745}{\includegraphics[scale=0.06]{orcid.pdf}\hspace{1mm}Nora M. Villanueva} \\
	Universidade de Vigo, Dep. of Statistics and  O.R. \\
    \& SiDOR Group, 36310 Vigo (Spain)\\
	\texttt{nmvillanueva@uvigo.gal} \\
	\And
	\href{https://orcid.org/0000-0003-4284-6509}{\includegraphics[scale=0.06]{orcid.pdf}\hspace{1mm}Marta Sestelo} \\
        Galician Centre for Mathematical Research \\
        and Technology (CITMAga),\\
        Santiago de Compostela (Spain)\\
        Universidade de Vigo, Dep. of Statistics and O.R. \\
        \& SiDOR Group, 36310 Vigo (Spain)\\
	\texttt{sestelo@uvigo.gal} \\}

\renewcommand{\shorttitle}{Cybersecurity threat detection based on a UEBA framework using Deep Autoencoders}

\hypersetup{
pdftitle={Cybersecurity threat detection based on a UEBA framework using Deep Autoencoders},
pdfsubject={stat.ME},
pdfauthor={Fuentes et al.},
pdfkeywords={First keyword, Second keyword, More},
}

\begin{document}

\vspace{3cm}

\bigskip

\bigskip

\bigskip

\bigskip

This article has been published in AIMS Mathematics, 2025, 10(10): 23496--23517. DOI: \href{https://www.aimspress.com/article/doi/10.3934/math.20251043}{10.3934/math.20251043}.

\newpage

\maketitle

\begin{abstract}
The increasing sophistication of cyberattacks, especially insider and process-related anomalies, poses a major challenge to enterprises, as traditional rule-based or shallow anomaly detection systems often fail to capture complex behavioural patterns. User and Entity Behaviour Analytics (UEBA) is a broad branch of data analytics that attempts to build a normal behavioural profile in order to detect anomalous events. Among the techniques used to detect anomalies, Deep Autoencoders constitute one of the most promising deep learning models on UEBA tasks, allowing explainable detection of security incidents that could lead to the leak of personal data, hijacking of systems, or access to sensitive business information. In this study, we introduce the first implementation of an explainable UEBA-based anomaly detection framework that leverages Deep Autoencoders in combination with Doc2Vec to process both numerical and textual features. Additionally, based on the theoretical foundations of neural networks, we offer a novel proof demonstrating the equivalence of two widely used definitions for fully-connected neural networks. The experimental results demonstrate the proposed framework's capability to detect real and synthetic anomalies effectively generated from real attack data, showing that the models provide not only correct identification of anomalies but also explainable results that enable the reconstruction of the possible origin of the anomaly. Compared to existing UEBA and anomaly detection approaches, the novelty of our framework lies in combining explainable multimodal feature processing with formal mathematical guarantees. Our findings suggest that the proposed UEBA framework can be seamlessly integrated into enterprise environments.
\end{abstract}


\maketitle

\section{Introduction}
\label{introduction}

In the current digital era, cybersecurity and the reliability of both physical and logical systems have become of increasing importance for industry and academia alike. The exponential growth of interconnected devices, the increasing volume of sensitive data, and the complexity of technological infrastructures highlight the need for robust algorithms that improve security and resilience. Mathematical models and methods play a central role in this task by offering formal frameworks to identify cyberattacks, develop cryptographic protocols, simulate potential incidents under a wide range of scenarios, and design defence strategies against adversarial threats. As these challenges intensify, mathematical approaches are becoming more crucial to guarantee robust system performance and to mitigate future cyber risks. Learning-based methods are increasingly employed to secure diverse environments. For example, in Internet of Things (IoT) systems, they are applied to detect attacks using heterogeneous sensor and network data \cite{Inayat2022}. In cyberphysical infrastructures, they have been explored to monitor and defend interconnected systems such as the energy-water nexus \cite{Khalid2020}. A related challenge arises in enterprise settings, where detecting anomalies in user activity ---such as signs of cyberattacks, rogue insiders, or negligent behaviour--- has become an essential task for modern security operations centres.

User and Entity Behaviour Analytics (UEBA) \cite{Shashanka_2016} is a powerful methodology to identify cyber threats by creating models of normal behavioural patterns and detecting deviations that may indicate malicious or negligent activities. To profile the behaviour of an entity, these models allow the incorporation of multiple sources of information such as sensor readings, network traffic, system logs, security alerts, email information, and even geo-positioned or biometric data. UEBA uses advanced statistical learning techniques to model the behaviour of users, employees, and customers, as well as machines, such as servers, switches, and personal systems. By analysing anomalies in the behaviour of users and devices, UEBA can detect intrusions, impersonation attacks, or negligent users \cite{Maher_2017}.

In parallel, Explainable Artificial Intelligence (xAI) has emerged as a key enabler in cybersecurity scenarios by addressing the growing need for interpretability and trust in complex AI-driven security systems. Since modern cybersecurity solutions rely on deep learning models, its inherent lack of transparency can prevent the analysts' ability to understand and respond to detected threats effectively. The integration of xAI techniques (such as SHAP, LIME or latent-space analysis) into cybersecurity operations enable cybersecurity experts to interpret and understand why a system has identified a specific event as anomalous behaviour, facilitating root-cause analysis and informed decision-making, prioritising genuine threats, and helping to identify false positives. Therefore, by integrating xAI techniques, UEBA-based cybersecurity tools not only improve technical performance, but also improve compliance with regulatory frameworks (such as the AI Act) that demand transparency in AI-based decision making, especially in sensitive areas such as finance. 

Several methods have been proposed in the literature related to the use of these techniques, which we summarise in Table \ref{table:related_methods}. In \cite{Shashanka_2016}, UEBA models based on Mahalanobis distance and Singular Value Decomposition (SVD) are implemented to identify anomalous behaviour in users accessing a server. Voris et al. \cite{Voris_2019} use Gaussian mixture models (GMM) for each computer,  collecting data for the file system, process launch, and network behaviour, in addition to establishing a series of trap files to attract and identify attackers. They also apply UEBA to continuously identify the user in the system by monitoring their activity. Another example is found in \cite{Pusara_2004}, where decision trees are applied to mouse movement data to identify the user. Similarly, in \cite{slipenchuk_epishkina_2019}, results of applying UEBA with mouse movement data, keyboard typing dynamics, and event sequences in the context of online banking operations are compared. In \cite{MENG20181}, a similar approach based on the Radial Basis Function Network (RBFN) classifier with Particle Swarm Optimisation (PSO) is applied to verify the user identity through touchscreen usage and other biometric data collected during web browsing. Moreover, the combination of static (logins, cookies, system type, etc.) and dynamic (mouse, keyboard, microphone, network usage, etc.) data to build user models was explored in \cite{Mart_n_2021}, where UEBA models are also used to guarantee user coherence when performing authentication with identity federations.

UEBA can be considered a use case for anomaly detection (or outlier detection) with personalised models. Once the users are identified, each model has to detect data points that do not conform to the expected behaviour. As anomaly detection has grown in popularity, a wide array of methods and techniques has emerged. Among these, autoencoders \cite{rumelhart_autoencoder} are a type of artificial neural network used for this purpose. Their goal is to learn (in an unsupervised way) a representation of the dataset by filtering out insignificant data or noise. Recent work in xAI for cybersecurity highlights the necessity of designing inherently interpretable (ante-hoc) models, prioritizing explainability principles from model conception through training \cite{ortega2024explainable}. A key advantage of the use of autoencoders for cybersecurity is that they tend to be more explainable than other deep learning models \cite{morales_2019case,GONZALEZMUNIZ2022108065}. Moreover, in cases where the Well-Defined Anomaly Distribution (WDAD) assumption does not hold \cite{phd10.5555/AAI27610132,2023dcnut}, autoencoders can be trained on data assumed to be normal (even if slightly contaminated) \cite{mauritz2021probabilistic}.

Autoencoders (AEs) have been used for anomaly detection since the work of Hawkins et al. \cite{Hawkins2002OutlierD}, and a general overview can be found in \cite{Wang_2019}. Examples include using Variational Autoencoders (VAEs) to construct anomaly scores \cite{xiao2020likelihood} and convolutional autoencoders for video signal anomaly detection \cite{Ribeiro_2018}. In \cite{Sakurada}, both incomplete and overcomplete autoencoders are applied in satellite data, while \cite{WANG2022110791} proposes a VAE coupled with a transformer architecture to account for dependencies in satellite data. In industrial anomaly detection, \cite{Zhou_2017} uses a norm-regularised autoencoder, and \cite{morales_2019case} combines an autoencoder with an LSTM network. Also, \cite{GONZALEZMUNIZ2022108065} utilises VAEs to classify anomalies in engineering systems. Finally, \cite{Ortega-Fernandez2023} showed that autoencoders outperform other algorithms in the detection of Denial of Service cyberattacks in industrial scenarios.

\begin{table}[ht!]
\caption{Comparison of related UEBA and anomaly-detection methods highlighting gaps addressed by our approach.}
\label{table:related_methods}
\begin{footnotesize}
\begin{center}
\begin{tabular}{p{5.0cm} p{1.5cm} p{7.5cm}}
\hline\noalign{\smallskip}
Method & Domain & Novelty / Differences w.r.t. this work \\
\hline\noalign{\smallskip}
Shashanka \textit{et al.} (2016) \cite{Shashanka_2016} & UEBA & Mahalanobis distance + SVD for user/device modelling; limited to numerical features without explainability; our work adds multimodal logs and interpretable residuals. \\
Voris \textit{et al.} (2019) \cite{Voris_2019} & UEBA & GMM with decoys for active authentication; trap-based, not scalable to enterprise monitoring; our work scales to large heterogeneous logs. \\
Pusara \& Brodley (2004) \cite{Pusara_2004} & Biometrics & Mouse-movement re-authentication; narrow biometric focus; our work targets full enterprise behaviour. \\
Slipenchuk \& Epishkina (2019) \cite{slipenchuk_epishkina_2019} & UEBA & Survey of UEBA statistical vs. ML methods; descriptive only, no unified DL framework; our work proposes a practical autoencoder pipeline. \\
Meng \textit{et al.} (2018) \cite{MENG20181} & Biometrics & Touch behaviour authentication with RBFN+PSO; mobile-focused, limited features; our work generalises to multimodal enterprise UEBA. \\
Mart\'in \textit{et al.} (2021) \cite{Mart_n_2021} & UEBA & UEBA for federated identity (OpenID Connect); restricted to identity scope; our work expands to enterprise-wide logs with textual features. \\
Morales-Forero \& Bassetto (2019) \cite{morales_2019case} & Industrial & Semi-supervised AE for Industrial Control Systems (ICS) anomaly detection; domain-specific, no text integration; our work extends AE to enterprise logs with multimodal features. \\
Gonz\'alez-Mu\~niz \textit{et al.} (2022) \cite{GONZALEZMUNIZ2022108065} & Industrial & VAE with two-step residual classification for engineering systems; limited to industrial datasets; our work adapts VAE to enterprise UEBA with textual features and explainability. \\
Sakurada \& Yairi (2014) \cite{Sakurada} & Satellite & AE for anomaly detection in satellite telemetry; domain-limited; our work applies AE to enterprise UEBA with multimodal features. \\
Wang \textit{et al.} (2022) \cite{WANG2022110791} & Satellite & Likelihood-based AE for telemetry anomaly detection; strong in satellite domain, not generalisable; our work extends to enterprise logs with interpretability. \\
Zhou \& Paffenroth (2017) \cite{Zhou_2017} & Industrial & Robust deep AE for anomaly detection; designed for ICS data; our work adapts robust AE concepts to UEBA with text+numeric logs. \\
Ortega-Fern\'andez \textit{et al.} (2023) \cite{Ortega-Fernandez2023} & Industrial & Deep AE network intrusion detection system designed to detect distributed denial-of-service (DDoS) attacks in ICS; effective but network-traffic only; our work generalises AE to enterprise UEBA with multimodal logs and interpretability. \\

\noalign{\smallskip}\hline
\end{tabular}
\end{center}
\end{footnotesize}
\end{table}

The motivation of this work lies in two key challenges of anomaly detection in enterprise environments. First, labelled attack data are rarely available, and the heterogeneity between infrastructures and behaviours limits the reuse of public datasets, which makes unsupervised approaches especially valuable. Second, most previous contributions focus only on numerical features, while enterprise logs also contain textual elements that are often ignored. In this work, we focus on executable paths as textual features, while noting that other sources, such as web addresses or email content, could also be explored in future research. In response, we propose a UEBA framework that integrates deep autoencoders with text embeddings, grounded on solid mathematical foundations.

Our work makes a twofold contribution. First, we provide novel theoretical results by proving the equivalence of two common definitions of fully-connected neural networks. Second, to the best of our knowledge, we present the first implementation of an explainable UEBA-based anomaly detection framework using autoencoders. Our methodology includes the use of text encoding models (Doc2Vec) alongside autoencoders to leverage both numerical and textual data, training  unlabelled (possibly contaminated) data, and using model residuals for explainability.  

This paper is structured as follows. Section~\ref{sec:methods} details the proposed methodology, including theoretical results and a description of the methods employed for feature extraction, residual space analysis, and the proposed architecture for UEBA-based anomaly detection. Section~\ref{sec:results} presents the results of the application of the proposed methodology to a real use-case of cybersecurity in a financial institution. Finally, Section~\ref{sec:conclusions} outlines the main conclusions and future research directions.

\section{Materials and methods}
\label{sec:methods}

In this work, we propose a novel UEBA-based anomaly detection framework based on Deep Autoencoders and the Doc2Vec algorithm for the pre-processing of text features. In the following Subsections \ref{sec:NN}-\ref{sec:tsne}, we describe the mathematical foundations of the used algorithms and present our theoretical contributions with a new proof of the equivalence of two common definitions of neural networks. Moreover, Subsection \ref{sec:implementation} describes the architectures of the UEBA-based anomaly detection framework. 

\subsection{Neural Networks}
\label{sec:NN}

Neural networks constitute a large set of learning models that originate from the early work on the Rosenblatt perceptron \cite{Rosenblatt_1958}. They approximate functions by interleaving affine transformations with non-linear activation functions. A feed-forward deep neural network uses longer chains of concatenated affine and activation functions to improve the representation of the target function. This definition of a feed-forward deep neural network can be formally expressed in the following manner.

\begin{definition}
\label{defNN}
Let $\mathcal{F}\subset \{\varphi:\mathbb{R}\to\mathbb{R}\}$ be a set of activation functions. Given $d\geq 2$ and an input dimension $n^{(0)}=n$, for each $l=1,\ldots,d$, let $n^{(l)}\in\mathbb{Z}^+$ with $n^{(d)}=m$, and let $A^{(l)}:\mathbb{R}^{n^{(l-1)}}\to\mathbb{R}^{n^{(l)}}$ be affine transformations. Define $\Phi^{(l)}:\mathbb{R}^{n^{(l)}}\to\mathbb{R}^{n^{(l)}}$ by
\[
\Phi^{(l)}(\mathbf{x})=(\varphi_1(x_1),\ldots,\varphi_{n^{(l)}}(x_{n^{(l)}})),
\]
with each $\varphi_j\in\mathcal{F}$ for $j=1,\ldots,n^{(l)}$. Then, a feed-forward deep neural network with $d-1$ hidden layers is the function
\[
\hat{f}=\Phi^{(d)}\circ A^{(d)}\circ \Phi^{(d-1)}\circ A^{(d-1)}\circ\cdots\circ\Phi^{(1)}\circ A^{(1)}:\mathbb{R}^{n}\to\mathbb{R}^{m}.
\]
\end{definition}

However, as noted in \cite{mhaskar2019function}, this definition does not uniquely determine the network structure and makes it difficult to formalise concepts such as sparsity and convolutions. An alternative, more constructive description is based on a layered graph where each node (or neuron) implements a simple function, as formalised by \cite{mhaskar2016deep,cano2017theory}. For brevity, we omit the definitions of layered graph and neuron, and they can be found in \cite{cano2017theory}. 

\begin{definition}
\label{defNN2}
Given a layered graph $\mathfrak{G}$, a feed-forward deep neural network with structure $\mathfrak{G}$ is any function defined on $\mathfrak{G}$ ($\mathfrak{G}$-function) such that each constituent function is a neuron.
\end{definition}

Neural networks of fixed depth or fixed width can approximate a wide range of functions modelling real-life processes \cite{Leshno_1993,kidger2020universal}. The universality property is crucial for any application; hence, it is significant to prove that both definitions are equivalent.
Below, we provide a mathematical proof (Proposition~\ref{prop_eq}) demonstrating this equivalence:

\begin{prop}
\label{prop_eq}
Definitions~\ref{defNN} and \ref{defNN2} are equivalent.
\end{prop}

\begin{proof}
We prove the equivalence of Definitions~\ref{defNN} and \ref{defNN2}.

\medskip

\noindent\textbf{(Definition~\ref{defNN} $\Rightarrow$ Definition~\ref{defNN2}):}  
Let $\hat{f}=\Phi^{(d)}\circ A^{(d)}\circ\cdots\circ\Phi^{(1)}\circ A^{(1)}$
with 
\[
\hat{f}^{(i)}=\Phi^{(i)}\circ A^{(i)}:\mathbb{R}^{n^{(i-1)}}\to\mathbb{R}^{n^{(i)}},\quad n^{(0)}=n,\; n^{(d)}=m.
\]
For $\mathbf{x}\in\mathbb{R}^n$, set
\[
\mathbf{h}^{(0)}=\mathbf{x},\quad \mathbf{h}^{(i)}=(\hat{f}^{(i)}\circ\cdots\circ\hat{f}^{(1)})(\mathbf{x}).
\]
Since each coordinate
\[
\pi_j\bigl(\hat{f}^{(i)}(\mathbf{h}^{(i-1)})\bigr)=\varphi_j\Big(W^{(i)}_j\cdot\mathbf{h}^{(i-1)}+b^{(i)}\Big)
\]
defines a neuron (i.e., the function $\pi_j\circ\hat{f}^{(i)}$), we construct a layered graph $\mathfrak{G}=(V,E)$ by:
\begin{enumerate}
    \item \emph{Input layer:} $V^{(0)}$ consists of $n$ nodes (the input coordinates \textbf{x}).
    \item \emph{Layers $1$ to $d$:} For each $i$, let $V^{(i)}$ consist of $n^{(i)}$ nodes, where each node $v_j\in V^{(i)}$ is assigned the function $\pi_j\circ\hat{f}^{(i)}$ and has incoming edges
    \[
    I_v^-=\{(u,v):\, u\in V^{(i-1)}\}.
    \]
\end{enumerate}
Thus, $\hat{f}$ is a $\mathfrak{G}$-function.

\medskip

\noindent\textbf{(Definition~\ref{defNN2} $\Rightarrow$ Definition~\ref{defNN}):}  
Conversely, let $\mathfrak{G}=(V,E)$ be a layered graph with layers
\[
V^{(i)}=\{v_1,\dots,v_{n^{(i)}}\}\quad (n^{(0)}=n,\; n^{(d)}=m),
\]
and let each node $v\in V^{(i)}$ have an associated function $f_v$. For $\mathbf{h}^{(i-1)}\in\mathbb{R}^{n^{(i-1)}}$, let $\mathbf{z}_v$ be the subvector of inputs corresponding to the predecessors of $v$. Define
\[
\hat{f}^{(i)}(\mathbf{h}^{(i-1)})=\bigl(f_{v_1}(\mathbf{z}_{v_1}),\dots,f_{v_{n^{(i)}}}(\mathbf{z}_{v_{n^{(i)}}})\bigr).
\]
Then, the overall network can be written as $
\hat{f}=\hat{f}^{(d)}\circ\cdots\circ\hat{f}^{(1)}$,
which is of the form given in Definition~\ref{defNN}.

\end{proof}

\subsection{Autoencoders}
\label{sec:auto}

An autoencoder is a model that approximates the identity function under a constraint that forces the model to capture the most salient features of the input. In our case, for a random sample $X=(\mathbf{x}_1,\ldots,\mathbf{x}_m)$ with $\mathbf{x}_i\in\mathbb{R}^n$, we define an autoencoder $\mathbf{AE}_n^p$ as follows.

\begin{definition}
Given positive integers $n$ and $p$ (with $p<n$), an autoencoder $\mathbf{AE}_n^p$ is a tuple
\[
(n, p, f, g, \mathcal{E}, \mathcal{D}, X, \Delta)
\]
where:
\begin{itemize}
    \item $\mathcal{E}$ and $\mathcal{D}$ are sets of functions from $\mathbb{R}^n$ to $\mathbb{R}^p$ and from $\mathbb{R}^p$ to $\mathbb{R}^n$, respectively;
    \item $f\in\mathcal{E}$ is the encoder and $g\in\mathcal{D}$ is the decoder;
    \item $\Delta$ is a dissimilarity measure (typically a metric) on $\mathbb{R}^n$.
\end{itemize}
\end{definition}

The latent space is the codomain of $f$, where the compressed representation of $\mathbf{x}$ is stored. The reconstruction error is defined as:

\begin{definition}
\label{rec}
The reconstruction error of the autoencoder is given by
\[
E_{f,g}(X)=\sum_{i=1}^m\Delta(g(f(\mathbf{x}_i)),\mathbf{x}_i).
\]
\end{definition}

Training an autoencoder involves finding functions $f$ and $g$ that minimise $E_{f,g}(X)$. In our work, both encoder and decoder are implemented as fully connected (regularised) neural networks. We focus on under-complete autoencoders ($p<n$) to force a compressed representation, which in turn leads the model to learn the dominant patterns of normal behaviour. Since anomalous data points are rare, the model prioritises the reconstruction of normal samples, making the reconstruction error an effective anomaly score.

It should be highlighted that Proposition~\ref{prop_eq} establishes that the two common formalisations of feed-forward neural networks (functional composition vs.\ layered graphs) are equivalent. Since our encoder and decoder are implemented as fully connected networks, this equivalence guarantees that the autoencoder used in our UEBA pipeline is a well-defined object in either formalism, with no loss of generality when moving between them.
This allows us to apply the classical universal approximation results for feed-forward networks \cite{Leshno_1993,kidger2020universal}, providing a rigorous foundation for our modelling choice: the autoencoder has sufficient expressivity to capture the dominant patterns of normal behaviour, while its residuals define a mathematically coherent and explainable anomaly score used throughout the framework.

\subsection{Doc2Vec}
\label{sec:doc2vec}

To process text-based variables (e.g., lists of executed processes), we use the Doc2Vec model \cite{quoc}. Doc2Vec is a neural network-based embedding method that learns vector representations of documents in an unsupervised manner. Similar to Word2Vec \cite{mikolov}, it clusters similar texts in the vector space. In the same way that Word2Vec embeds related words like synonyms or topics close in the vector space, Doc2Vec also clusters similar texts together, for example, by identifying the topic of the text or by finding similar words between texts. Two main algorithms exist for training Doc2Vec: Distributed Bag of Words (DBOW) and Distributed Memory (DM). In DBOW, the document vector is used to predict random word vectors from the document; in DM, both the document vector and word vectors are used to predict the next word in a sequence. The resulting document embeddings capture semantic similarities that are later used in our UEBA framework by applying the Doc2Vec trained with DBOW to extract information from the lists of processes recovered from the activity logs. We specifically adopt Doc2Vec with the DBOW algorithm because executable paths behave more like structured categorical tokens than natural language, so co-occurrence at the window level captures useful behavioural similarity (e.g., programs launched together) without requiring heavy language models.

\subsection{t-distributed Stochastic Neighbour Embedding} \label{sec:tsne}

The t-distributed Stochastic Neighbour Embedding (t-SNE) \cite{van2008visualizing} is a dimensionality reduction technique used primarily for visualising high-dimensional data in two or three dimensions. It preserves local structures by mapping similar points from high-dimensional space to nearby points in the lower-dimensional embedding, while distant points remain separated. t-SNE works by first computing pairwise conditional probabilities based on distances between points in the high-dimensional space, and modelling the local similarities as Gaussian distributions. It then maps these points into a lower-dimensional space using a Student's t-distribution to model the similarity of the embeddings. The algorithm iteratively adjusts embedding positions by minimising the Kullback-Leibler divergence between these two distributions:
\begin{equation*}
\text{KL}(P||Q) = \sum_{i \neq j} p_{ij}\log\frac{p_{ij}}{q_{ij}},
\end{equation*}
where $P$ and $Q$ represent the joint probability distributions of pairwise similarities between data points in the high-dimensional and low-dimensional spaces, respectively. This lower dimensionality embedding allows us to identify patterns that are characterised by their local structure, such as clusters and anomalies in high dimensional datasets.

We use t-SNE to analyse the dataset and validate the autoencoder model's behaviour. We apply t-SNE to the test data to analyse the presence of clusters indicative of different user groups and behaviours, and validate the feasibility of a UEBA-based approach. Moreover, we also use t-SNE to study the residuals of the reconstruction error of the model, verifying that anomalous data points remain distinguishable within the residual space.

\subsection{UEBA-based anomaly detection framework}
\label{sec:implementation}

The dataset used in this study is derived from multiple real data sources, including Windows events of user activity, emails, and antivirus logs from a financial institution. For this reason, preprocessing plays a crucial role in preparing the raw data for effective anomaly detection. The preprocessing steps involve cleaning, transforming, and encoding the data using the Doc2Vec model described in Section~\ref{sec:doc2vec} to derive features into a format suitable for the anomaly detection model. 

The data is ingested as time-series from logs and is aggregated into fixed windows, summarising them into key statistics, like total counts and average time intervals. In addition, we perform feature engineering to derive new variables that better capture behavioural patterns, including metrics such as the average time between logins, the ratio of failed to successful logins, and the frequency of antivirus alerts. The set of derived features is detailed in Table~\ref{table:features}, with 2 indexing variables and 19 features.

For handling missing values, a lack of activity is assumed to correspond to a zero count, while timing variables (e.g., \texttt{avg\_sec\_bet\_logins}) are imputed using the duration of the aggregation window in seconds as a maximum time. Meanwhile, text data (specifically, the executable names from processes executed within each window) are combined into a single field (\texttt{process\_list}) and encoded using a Doc2Vec model (trained with DBOW) to generate a 64-dimensional embedding vector for each window. These embeddings are concatenated to the derived numerical features, obtaining the final input vector with 83 features. 

Finally, all numerical features are normalised using a robust scaler followed by a min-max scaler to ensure that each variable contributes equally to the model. At the end of this preprocessing pipeline, the dataset is clean, structured, and ready to be used in the UEBA models and anomaly detection framework.

\begin{table}[h]
\caption{Variables collected in the dataset.}
\footnotesize
\label{table:features}
\begin{footnotesize}
\begin{center}
\begin{tabular}{p{3cm}p{12cm}p{1cm}}
\hline\noalign{\smallskip}
Variable & Description & Type \\ 
\hline\noalign{\smallskip}
\texttt{time} & Date and time when the data was generated (used for window aggregation and indexing only) & date \\ 
\texttt{CallerUser} & Username (used for role aggregation and indexing only). & factor \\ 
\texttt{WorkstationName} & Machine name where the data was generated. & factor \\ 
\texttt{num\_new\_process} & Number of new processes created (e.g., Windows event 4688). & numeric \\ 
\texttt{num\_logins} & Number of successful logins (e.g., Windows event 4624). & numeric \\ 
\texttt{avg\_sec\_bet\_logins} & Average time in seconds between successful logins. & numeric \\ 
\texttt{num\_f\_logins} & Number of failed login attempts (e.g., Windows event 4625). & numeric \\ 
\texttt{avg\_sec\_bet\_f\_logins} & Average time between failed logins. & numeric \\ 
\texttt{num\_antivirus\_alerts} & Number of incidents detected by the antivirus. & numeric \\ 
\texttt{num\_firewall\_alerts} & Number of incidents detected by the firewall. & numeric \\ 
\texttt{sent\_emails} & Number of emails sent by the user. & numeric \\ 
\texttt{received\_emails} & Number of emails received by the user. & numeric \\ 
\texttt{incident\_emails} & Number of emails flagged as incidents. & numeric \\ 
\texttt{sent\_emails\_size} & Total size of sent emails (body and attachments). & numeric \\ 
\texttt{received\_emails\_size} & Total size of received emails (body and attachments). & numeric \\ 
\texttt{sent\_email\_files} & Number of file attachments in sent emails. & numeric \\ 
\texttt{received\_email\_files} & Number of file attachments in received emails. & numeric \\ 
\texttt{sent\_email\_links} & Number of web links in sent emails. & numeric \\ 
\texttt{received\_email\_links} & Number of web links in received emails. & numeric \\ 
\texttt{4100\_events}  & Number of PowerShell errors (e.g., Windows event 4100). & numeric \\ 
\texttt{4104\_events} & Number of remote PowerShell commands executed (e.g., Windows event 4104). & numeric \\ 

\noalign{\smallskip}\hline
\end{tabular}
\end{center}
\end{footnotesize}
\end{table}

\begin{figure}[h]
\centering
\includegraphics[width=\textwidth]{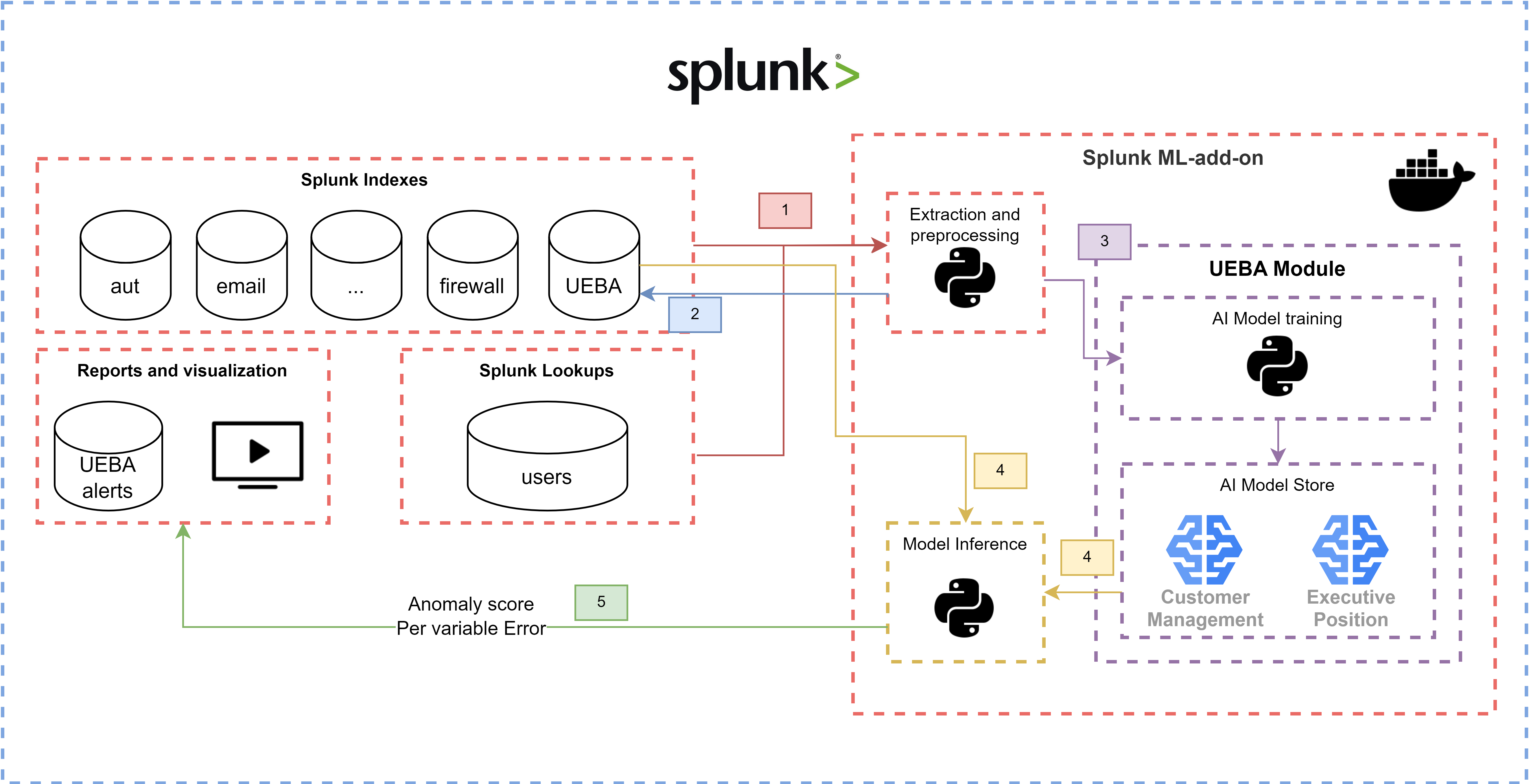}
\caption{Architecture of the UEBA-based anomaly detector.}
\label{fig:pipeline}
\end{figure}

\begin{figure}[h]
\centering
\includegraphics[width=\textwidth]{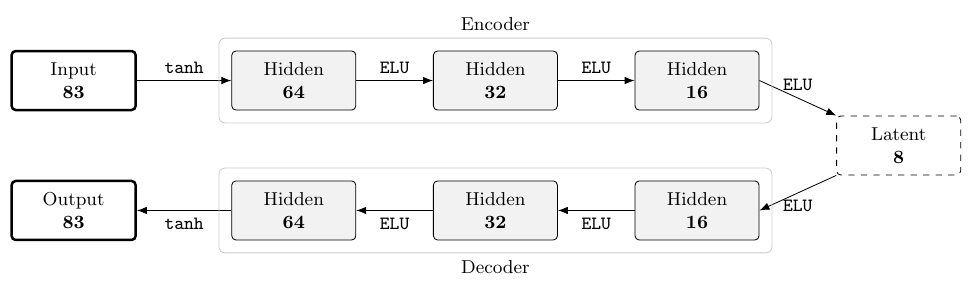}
\caption{Network diagram of the proposed autoencoder.}
\label{fig:network}
\end{figure}

Figure~\ref{fig:pipeline} illustrates the overall architecture of the proposed UEBA-based anomaly detection framework, where the numbered arrows correspond to the different stages of the process (1--5). The process starts with data collection (1) from multiple sources (e.g., Windows events, emails, antivirus, firewall), which are stored in Splunk Enterprise, a common choice in the industry to collect, index, search, analyse, and visualise large volumes of machine-generated data in real time, as time-series events. The feature extraction pipeline aggregates and processes these events into summary variables, while the text data is encoded using Doc2Vec (see Table~\ref{table:features}). Once the raw data has been transformed into security events and the features have been computed, this aggregated data  is grouped into entities based on business roles, e.g. Customer  (CM) and Executive Positions (EP)  and stored at the UEBA index in Splunk (2). This aggregated UEBA dataset can now be used to train the Autoencoder and Doc2Vec models (see Algorithm~\ref{alg:training}), which are later stored in a model store (3). The hyperparameters used to train both models are available in Table \ref{table:hyperparams}, and the network diagram is presented in Figure \ref{fig:network}. Once the models are trained and stored, they are made available through the model store (4) to generate predictions for new data (see Algorithm~\ref{alg:inference}) coming from the index (4) in near real-time, including the residual-based anomaly scores which are stored in the alert databased, and therefore available for plotting and advanced analysis (5).

\begin{table}[h]
\centering
\small
\caption{Hyperparameters of the UEBA anomaly detection framework for the Customer Management (CM) and Executive Positions(EP) roles}
\label{table:hyperparams}
\begin{tabular}{p{3cm} p{10cm}}
\hline\noalign{\smallskip}
\textbf{Component} & \textbf{Setting} \\
\hline\noalign{\smallskip}
Doc2Vec & 64-dimensional embeddings; DBOW architecture; window size = 5; epochs = 20 \\
Autoencoder & Input size = 83 features; hidden layers = [64, 32, 16, 8, 16 32, 64]; latent dimension = 8; activations = ELU (internal), \texttt{tanh} (first/last) \\
Training & Optimiser = Adam (lr = 0.001 (CM) / 0.01 (EP)\tablefootnote{The higher learning rate in the EP role was selected via grid search in combination with the larger batch size, resulting in stable convergence and improved generalisation.}); batch size = 64 (CM) / 256 (EP); early stopping (patience = 10, monitor = validation MSE); $L_1$ regularization ($\lambda=0.001$); validation split = 20\% \\
\hline
\end{tabular}
\end{table}

\begin{algorithm}[H]
\caption{Training of the UEBA framework}
\label{alg:training}
\KwIn{Windowed log events $\{\mathcal{E}_i\}$}
\KwOut{Trained autoencoder $f_{\theta}$, Doc2Vec encoder $\phi$, scaler, decision threshold $\tau$}
\begin{enumerate}
  \item Extract numerical features $n_i=g_{\text{num}}(\mathcal{E}_i)$.
  \item Encode process list tokens $\mathcal{T}_i$ with Doc2Vec (DBOW) to obtain $t_i=\phi(\mathcal{T}_i)$.
  \item Concatenate: $x_i=[n_i;t_i]$, then apply scaling to get $\tilde{x}_i$.
  \item Train deep autoencoder $f_{\theta}$ to minimize reconstruction loss with $L_1$ regularization and early stopping.
  \item Compute scores $s_i=\|\tilde{x}_i-f_{\theta}(\tilde{x}_i)\|_{1}$ on validation data.
  \item Set threshold $\tau$ as the 95th percentile of $\{s_i\}$.
\end{enumerate}
\end{algorithm}

\begin{algorithm}[H]
\caption{Inference and anomaly scoring}
\label{alg:inference}
\KwIn{New window $\mathcal{E}$; trained $f_{\theta}$, $\phi$, scaler; threshold $\tau$}
\KwOut{Score $s$, residual vector $r$, decision $y\in\{\text{normal},\text{anomaly}\}$}
\begin{enumerate}
  \item Extract features: $n=g_{\text{num}}(\mathcal{E})$, $t=\phi(\mathcal{T})$, $x=[n;t]$.
  \item Scale input: $\tilde{x}=\mathrm{scale}(x)$.
  \item Reconstruct: $\hat{x}=f_{\theta}(\tilde{x})$.
  \item Compute residuals $r=\tilde{x}-\hat{x}$ and anomaly score $s=\|r\|_{1}$.
  \item Decision: $y=\mathbb{I}[s\geq\tau]$. Return $(s,r,y)$ for alerting and visualisation.
\end{enumerate}
\end{algorithm}

We generated two different datasets of one year of historical data, one for the Customer Management group (25313 records) and the other for the Executive Positions group (9804 records). Both datasets contain records of user behaviour that have been cleared as normal behaviour by the existing security filters from the institution; however, they may contain a small number of anomalies that evaded these measures. For this reason, we assume the data to be contaminated data \cite{tian2023leveraging}, and thus unlabelled. Both datasets are split into training, validation, and testing sets using a standard split: 20\% of the records are held out as the testing set, and the remaining 80\% is used for training, of which 20\% is further reserved as a validation set. We train a separate deep autoencoder model for each of the user groups. The model architecture starts with 83 input features (including the 19 aggregated features and the 64 components of the Doc2Vec transformation) and compresses them to an 8-dimensional latent space via three hidden layers (with 64, 32, and 16 neurons, respectively). The encoder and decoder networks use ELU activations (except in the first and last layers, which use \texttt{tanh}). Training employs the Adam optimiser, early stopping, and $L_1$ regularisation. A decision threshold $\tau$  is determined as the 95th percentile of the reconstruction error on the validation set. 

Trained models are stored using MLflow and later deployed to analyse incoming data. The anomaly score (based on the reconstruction error) and auxiliary statistics are sent back to Splunk for reporting and further analysis.

\section{Results and Discussion}
\label{sec:results}

We evaluate the proposed UEBA framework in two complementary settings. First, we conduct experiments on real-world data collected from a financial institution in Section \ref{sec:results-real}, providing a strong validation of the proposed framework under real operational conditions. Next, we extend these findings in Section \ref{sec:results-sim} using simulated data sampled from real attack scenarios, allowing us to systematically probe the system's response to specific threat vectors and anomalous behaviour of different intensities. By combining these two perspectives--actual enterprise data and controlled simulations—our analysis offers a robust demonstration of how UEBA can enhance security monitoring, highlight anomalous user or entity behaviour, and detect sophisticated cyber threats in an explainable manner.

In the following sections, we present the results of these evaluations, focusing on detection rate, residual-space analysis using t-SNE projections, performance under synthetic anomalies, and overall explainability.

\subsection{Performance on real data} \label{sec:results-real}

Evaluating unsupervised models is challenging in the absence of labelled data in real scenarios. To evaluate the anomaly detection capabilities of the proposed framework, we assess its ability to learn normal behavioural patterns by analysing the positive rate on the test set. The decision threshold $\tau$ is fixed at the 95th percentile of the reconstruction error on the validation set.  
Table~\ref{table:positiverate} shows the positive rates for both user groups (Customer Management and Executive Positions). We can observe how positive rates for both groups are close to the 5\% value, indicating that the models are well-calibrated. 
While the overall calibration is consistent across groups, 
we note that the positive rate of the Executive Positions model 
(4.61\%) is slightly below the 5\% target, whereas the Customer 
Management model (5.09\%) is slightly above it. This difference is 
minor in absolute terms, but it reflects the fact that user groups 
with fewer samples and more heterogeneous activity patterns may lead to tighter thresholds and a slightly more conservative model. Such variations are expected in UEBA applications, where behaviour differs by role, and they showcase the importance of tailoring models to specific roles or groups.

These positive-rate results constitute the primary evaluation of our framework in an unsupervised setting. 
Since the threshold was fixed during training, the fact that calibration holds on the test set shows that the models generalise well without relying on labelled anomalies. Standard supervised metrics such as precision, recall, or F1 cannot be meaningfully computed here, since the dataset lacks anomaly labels, and the data is taken as possibly contaminated data. In addition, as reported in Section~\ref{sec:results-sim}, we were provided with a small set of attack events.  Although these real anomalies were valuable, their very limited number made the evaluation less informative: the models detected them almost trivially, with a true positive rate close to $100\%$. This limitation motivated the use of synthetic anomaly experiments, which should be regarded as a complementary stress test probing robustness under varying anomaly intensity rather than as the main evaluation.

\begin{figure}[ht]
\begin{subfigure}{0.48\linewidth}
  \centering
  \includegraphics[width=\linewidth]{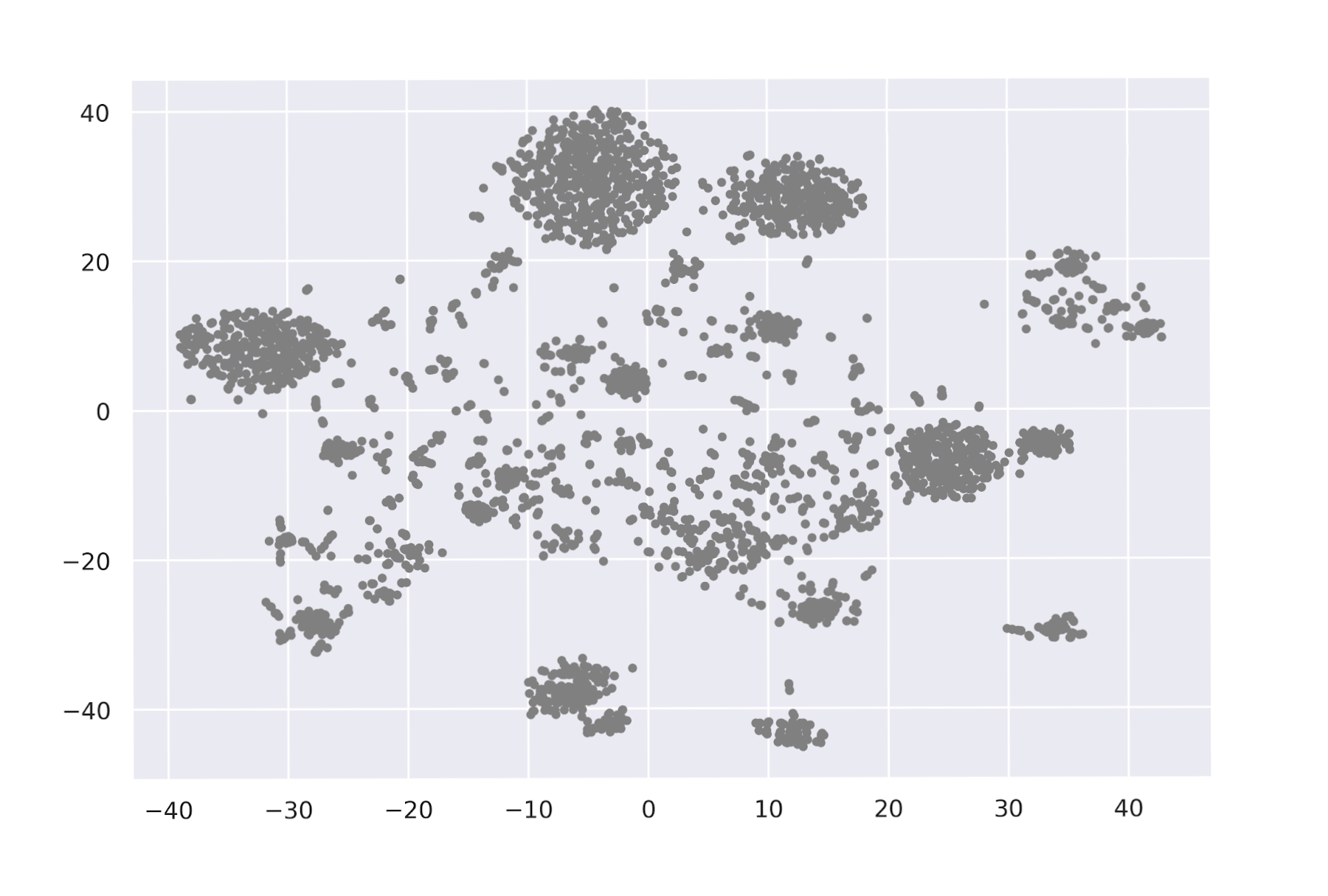}
  \caption{Normal data.}
\end{subfigure}
\hfill
\begin{subfigure}{0.48\linewidth}
  \centering
  \includegraphics[width=\linewidth]{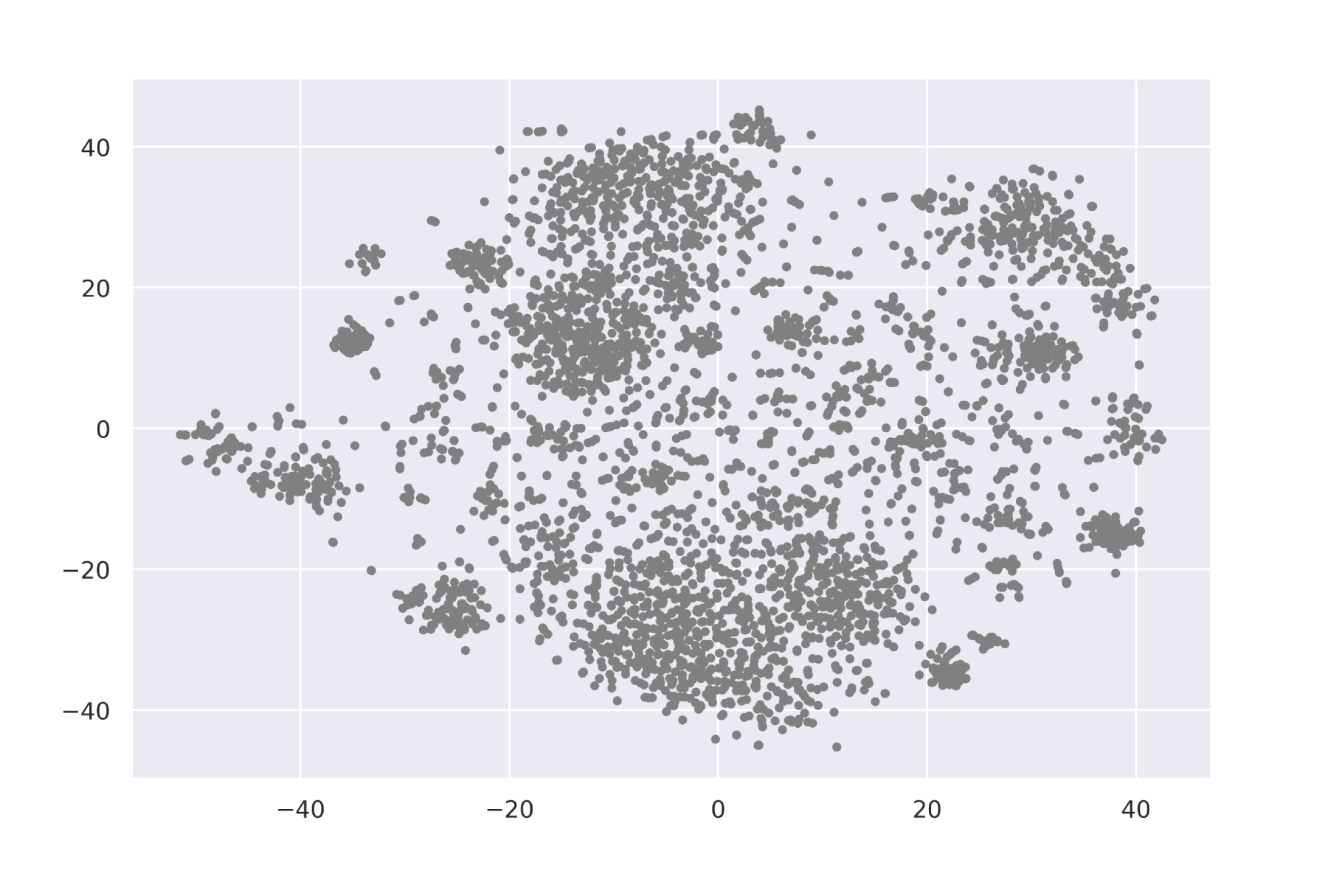}
  \caption{Residuals.}
\end{subfigure}
\caption{t-SNE representation of Customer Management. (a) Normal test data and (b) corresponding residuals.}
\label{fig:fig_datares}
\end{figure}

Moreover, to better understand the model's learning effectiveness, we perform a t-SNE projection of both the original test data and the corresponding residuals. Figure~\ref{fig:fig_datares} shows a t-SNE projection of normal test data from the Customer Management group (left) and the corresponding residuals (right). We can observe how the residual space shows less dependency clustering compared to the original data, confirming that the model has successfully captured the dominant patterns of normal behaviour. The residual distribution reveals anomalies as scattered points on the edges.

\begin{table}[h]
\centering
\caption{Positive rates for the two autoencoder models.}
\label{table:positiverate}
\begin{tabular}{l c}
\hline\noalign{\smallskip}
UEBA Model & Positive Rate \\
\hline\noalign{\smallskip}
Customer Management  & 5.09\% \\
Executive Positions & 4.61\% \\
\noalign{\smallskip}\hline
\end{tabular}
\end{table}

\FloatBarrier

\break

\subsection{Performance on synthetic anomalies} \label{sec:results-sim}

To assess the model's capability in detecting real-world anomalies, we conducted an experiment with synthetic anomalies generated from 10 real attack scenarios provided by the financial institution (login anomalies, antivirus incidents, email anomalies, and process-related anomalies). These synthetic anomalies are generated by taking convex combinations of the real anomalies with normal behaviour data. This procedure allows us to increase the sample size of the test set, using the variability of normal behaviour to provide more varied anomalies and study the model's detection capability as a function of an anomaly intensity factor, $\lambda_k$. Particularly, for each $j = 1, \ldots, 10$,  and for $k = 1, \ldots, 100$, we obtain a synthetic test set as follows
\[
\mathbf{a}_k^{*j}=\mathbf{z}_j(1-\lambda_k)+\lambda_k \mathbf{a}_j,
\]
\noindent where $\mathbf{z}_j$ is a randomly sampled element with normal behaviour data from the test set, $\lambda_k \in \left[ 0,1 \right]$ is the anomaly intensity factor, which takes values in steps of 0.01, and $\mathbf{a}_j$ is a real-type anomaly. Note that, with this procedure, we obtain a synthetic test set of sample size 1000. 

This convex interpolation scheme provides a simple yet controlled way to validate the anomaly detection sensitivity. 
By varying the anomaly intensity factor $\lambda$, we can gradually shift normal samples toward real anomalies and observe how the model responds, while ensuring that the resulting points remain inside the convex hull of the observed feature space, reducing the risk of generating unrealistic samples. 
At the same time, this construction mainly alters feature magnitudes and does not capture temporal dependencies, cross-feature correlations, or coordinated attack strategies. 
Therefore, synthetic anomalies should be regarded as a complementary stress test of robustness, rather than a substitute for larger-scale evaluation on real-world attack data.

\newpage

\begin{figure}[h!]
\centering
  \includegraphics[width=\linewidth]{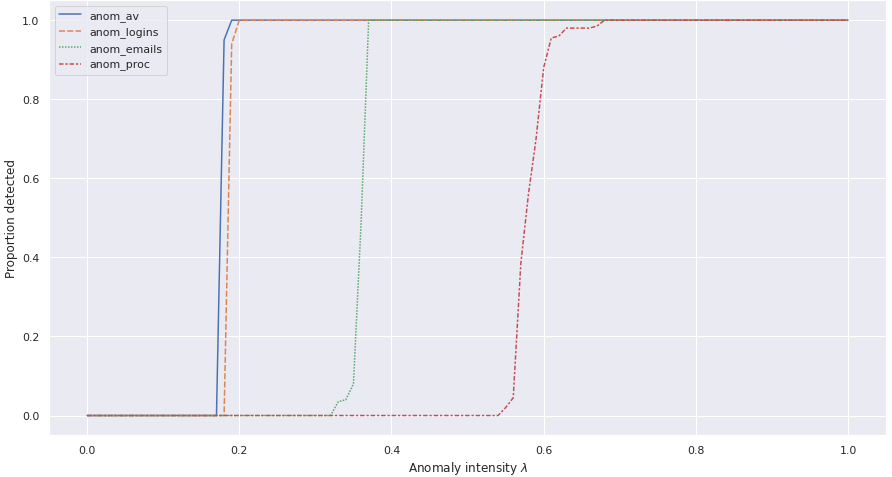}
  \caption{Anomaly detection rates as a function of anomaly intensity for each model for the Customer Management model.}
\label{fig:fig_detection1}
\end{figure}

\begin{figure}[h!]
\centering
  \includegraphics[width=\linewidth]{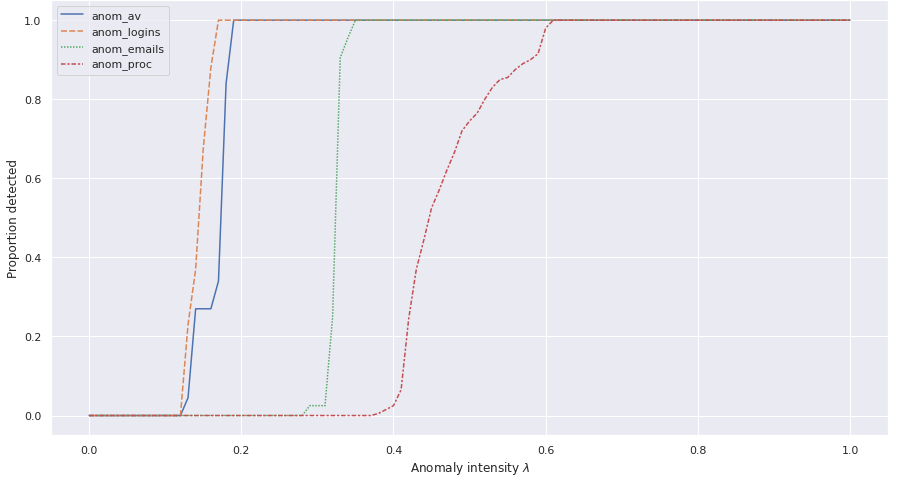}
  \caption{Anomaly detection rates as a function of anomaly intensity for each model for the Executive Positions model.}
\label{fig:fig_detection2}
\end{figure}
\FloatBarrier

\newpage

Figure~\ref{fig:fig_detection1} and \ref{fig:fig_detection2} present the detection rate as a function of the anomaly intensity factor $\lambda$ for both user groups (Customer Management and Executive Positions). The results demonstrate that the models reliably detect anomalies when $\lambda> 0.7$ for all anomaly types. For specific types, such as login and antivirus anomalies, detection occurs at much lower intensity levels ($\lambda > 0.2$). In contrast, process anomalies require higher intensity levels for reliable detection, primarily due to the complexity of encoding text-based data. While Doc2Vec provides a compact and efficient encoding, its averaging nature can obscure rare but highly informative tokens. More expressive embeddings could, in principle, capture richer temporal and syntactic structure, though at higher computational cost (see Section \ref{sec:conclusions} for possible alternatives for future work). 

These detection-vs-intensity curves serve as surrogate evaluation metrics in the absence of extensive labelled examples. Rather than relying on 
precision or recall, which cannot be meaningfully computed here, we evaluate how detection rates evolve as normal samples are gradually 
shifted toward known attack behaviours. This provides a principled and interpretable way to assess sensitivity under scarce-label conditions, 
showing that the framework can reliably separate normal from anomalous behaviour once the anomalies reach sufficient intensity. In this sense, 
the detection--vs--intensity curves can be interpreted as an analogue of statistical power curves: the anomaly intensity parameter $\lambda$ plays 
a role similar to the effect of the sample size, and the resulting curves quantify how quickly the detector achieves high detection probability as anomalies become 
more pronounced.

\begin{figure}[ht]
\centering
\begin{subfigure}{0.9\textwidth}
  \centering
  \includegraphics[width=0.9\textwidth]{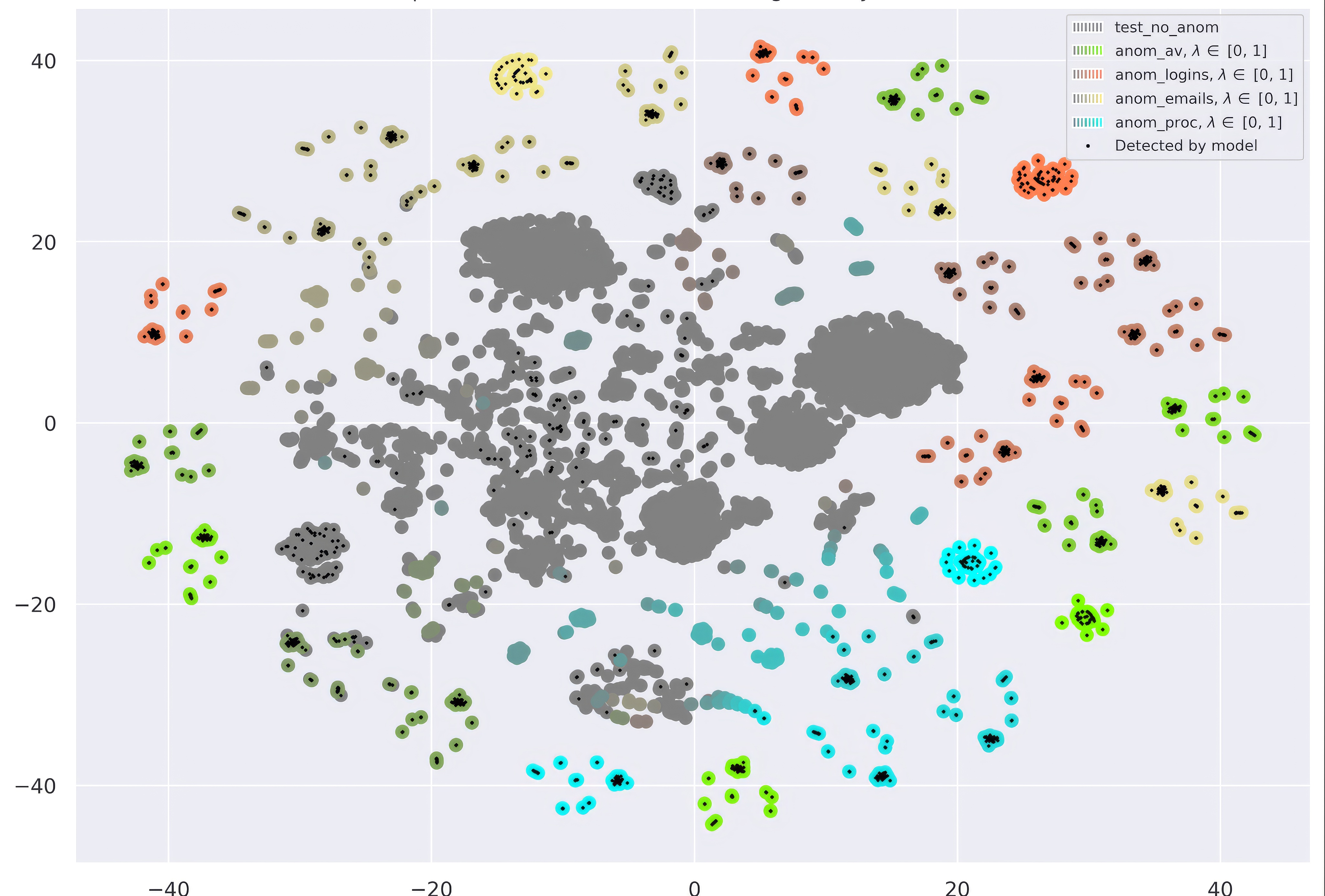}
  \caption{Sample of test data with anomalies.}
  \label{fig:sfig21}
\end{subfigure}\\[1mm]
\begin{subfigure}{0.9\textwidth}
  \centering
  \includegraphics[width=0.9\textwidth]{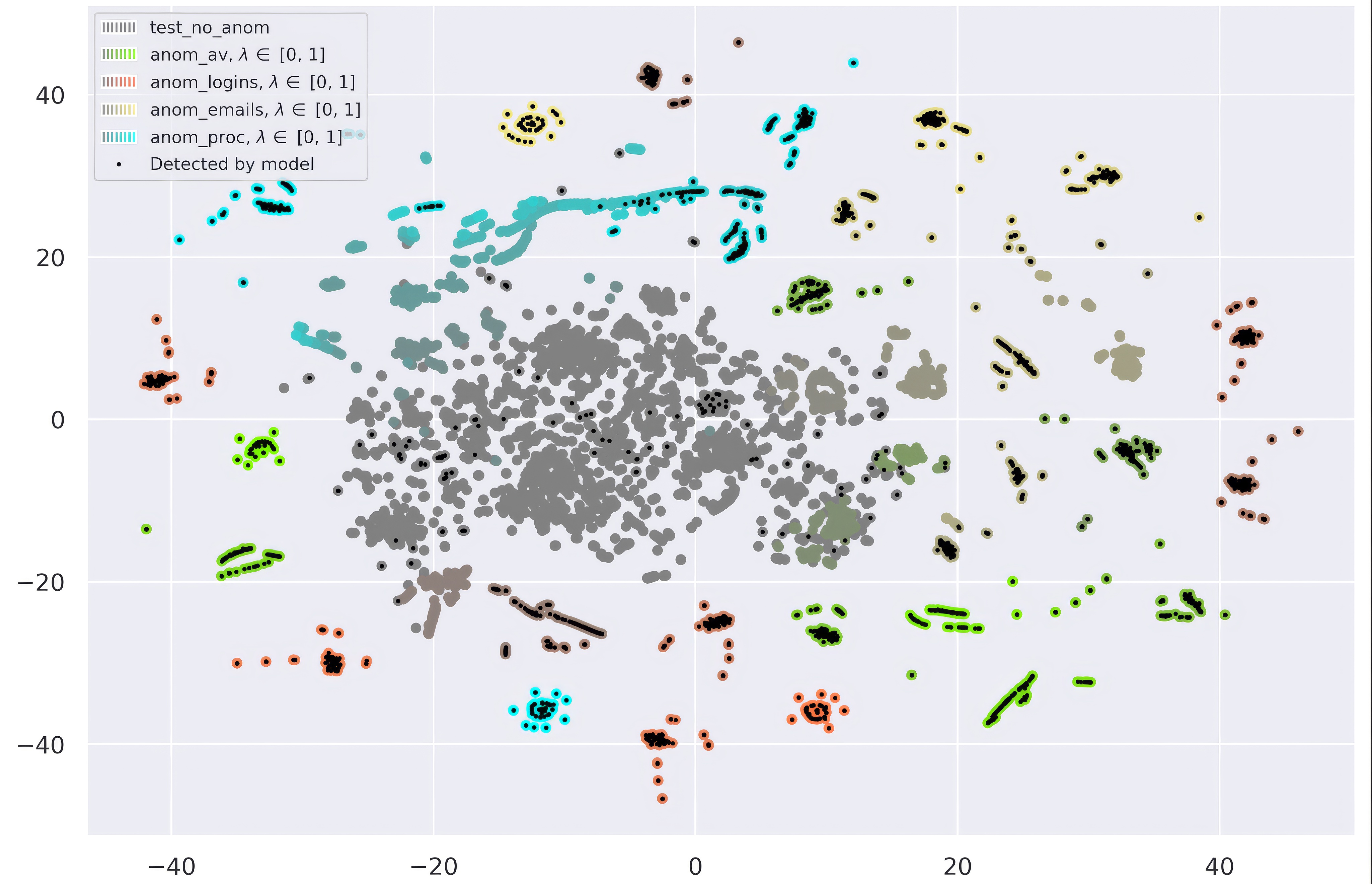}
  \caption{Corresponding residuals.}
  \label{fig:sfig22}
\end{subfigure}
\caption{t-SNE representations of data and residuals for the Customer Management group. Dotted points indicate instances flagged as anomalies. Saturation indicates the intensity of anomalies.}
\label{fig:fig_tsne}
\end{figure}

Figure~\ref{fig:fig_tsne} shows a t-SNE embedding of the test set. Each colour represents a different anomaly type (login, email, antivirus and process), with colour and saturation indicating the type and intensity of anomalies. The results confirm that anomalies become increasingly distinguishable as their intensity ($\lambda$) increases. Notably, anomalies with high $\lambda$ values form clear clusters in the residual space, confirming the model's capability to separate abnormal behaviour from normal patterns. Figure~\ref{fig:sfig21} illustrates how synthetic anomalies are embedded alongside normal data in the original feature space, while Figure~\ref{fig:sfig22} shows the corresponding residuals, where anomalies show clearer clusters, thus being easier to identify. Process-related anomalies appear less distinct due to their complexity and dependence on text-based feature encoding.

\begin{figure}[ht]
\centering
\begin{subfigure}{0.9\linewidth}
  \includegraphics[width=\linewidth]{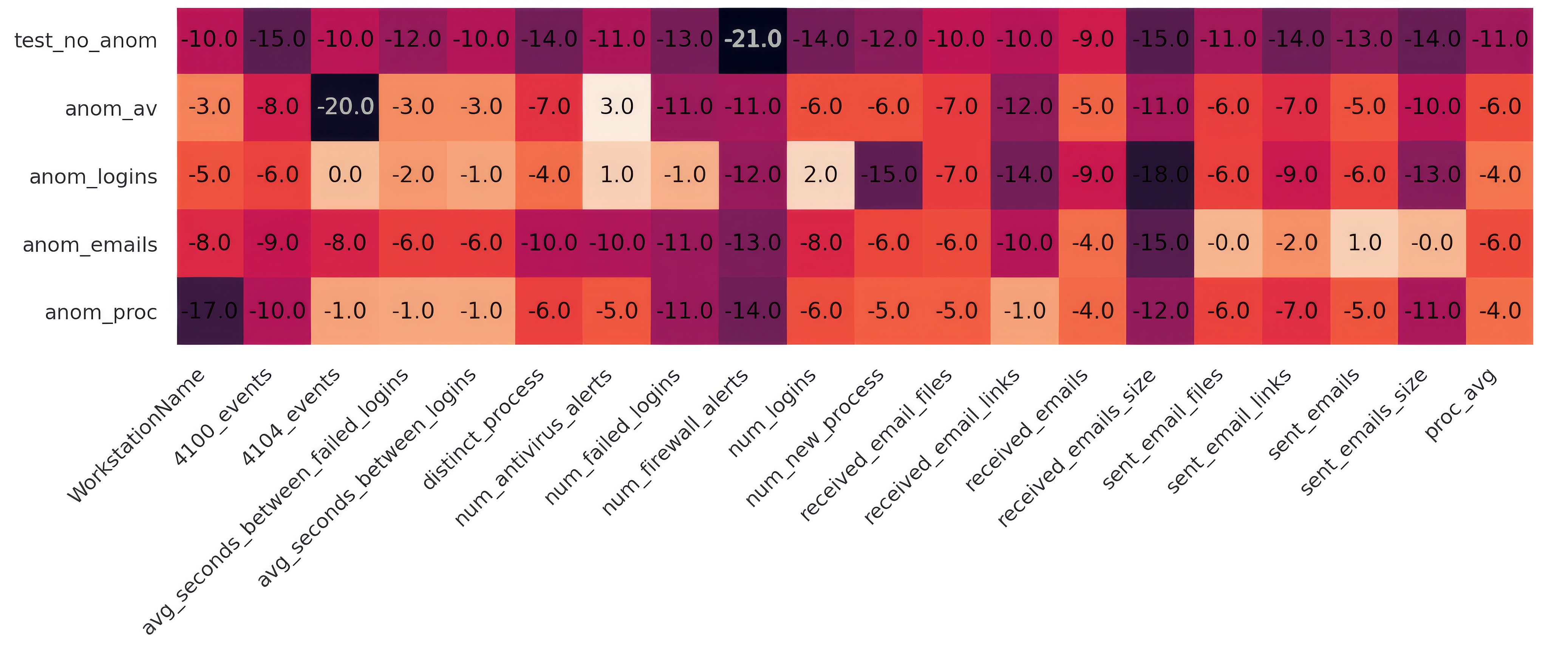}
  \caption{Customer Management.}
\end{subfigure}\\[2mm]
\begin{subfigure}{0.9\linewidth}
  \includegraphics[width=\linewidth]{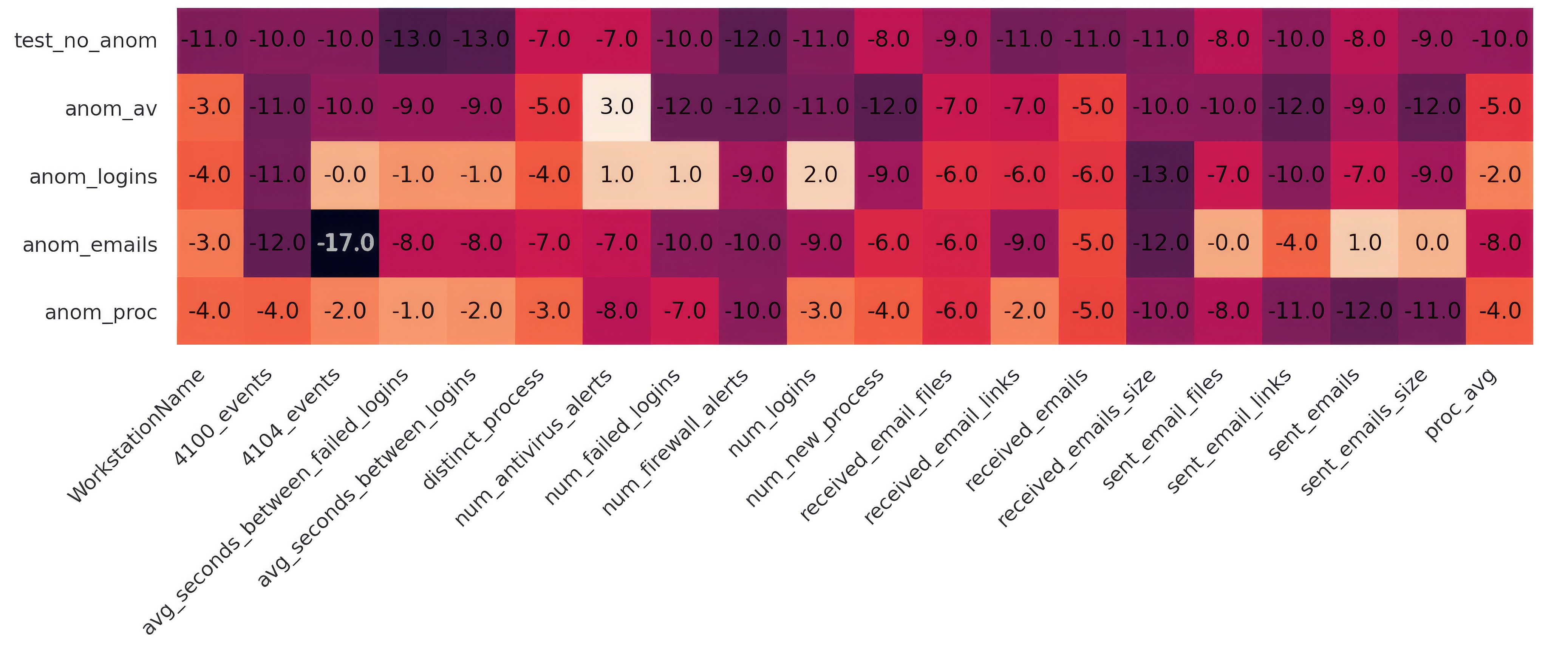}
  \caption{Executive Positions.}
\end{subfigure}
\caption{Logarithm of the reconstruction error per feature for each model.}
\label{fig:fig_var}
\end{figure}

At last, we will assess the proposed methodology's ability to provide explainable model results through the per-feature reconstruction error. Figure~\ref{fig:fig_var} shows the logarithm of the per-feature reconstruction error for each model for fully anomalous data ($\lambda=1$). We can observe that they are easier to detect, and this effect becomes more evident, with higher errors appearing on features related to the anomaly. For instance, the email-related anomalies show the highest reconstruction error on the \texttt{sent\_email\_*} variables. In addition, in the case of login anomalies, the highest errors are observed in the antivirus and login variables.

However, for anomalies that are more challenging, such as process anomalies, the per-variable reconstruction error alone may not suffice in identifying the origin of the anomaly. We speculate that the reason these anomalies are harder to detect and explain is that they heavily depend on the process encoding by Doc2Vec. However, we have to remark that, when removing the text variables, the detection performance worsened, so the text encoding provides valuable information for the detection of these anomalies, even if it\'s not enough for clear interpretation.

\vspace{1cm}
\FloatBarrier

\FloatBarrier

We can observe how features related to login patterns, email attachment size, and failed login attempts exhibit the highest contributions to the anomaly score, offering clear indications of potential security incidents. Anomalies in email activity, such as unusually large attachments or an abnormal volume of sent emails, are immediately recognisable. Process-related anomalies, on the other hand, show a more distributed error pattern due to the diverse and complex nature of process command sequences. 

These results highlight the strength of the proposed framework in detecting a wide range of anomalies with high accuracy while being explainable. The combination of Deep Autoencoders and Doc2Vec allows for an effective integration of numerical and text-based features, providing a comprehensive view of user behaviour. The ability to visualise residuals and analyse feature-level contributions significantly enhances explainability, making the framework more practical and effective for its use in cybersecurity applications. 

\section{Conclusions}
\label{sec:conclusions}

This study presents a UEBA-based anomaly detection framework that leverages Deep Autoencoders to identify suspicious activities within a real-world cybersecurity use case in a financial institution. By integrating both numerical and text-based features through Doc2Vec embeddings, the proposed approach can capture complex behavioural patterns to detect anomalies that may otherwise remain undetected by more traditional methods. A key contribution is a novel theoretical result proving the equivalence of two common definitions of fully connected neural networks, thereby grounding our autoencoder design within universal approximation theory. In addition, we provide a set of experimental evaluations to showcase how advanced deep learning techniques can be employed for explainable, behaviour-based anomaly detection.

Experimental evaluations showed that the proposed anomaly detection framework achieves a high detection rate, even in challenging conditions with contaminated training data. Additional experiments with synthetic anomalies--reflecting anomalous login patterns, email activity anomalies, and antivirus alerts--confirmed the framework's robustness and adaptability to diverse cybersecurity threats. A key advantage of our method lies in the ability to perform a residual-based analysis, which enhances explainability by pinpointing specific features that deviate from the normal baseline profile. This explainability is essential in practical cybersecurity contexts where timely responses and a deep understanding of anomalies can significantly improve investigation efficiency and reduce false alarms.

Compared to more traditional anomaly detection techniques such as PCA-based monitoring, statistical thresholds, or clustering approaches, our framework offers several distinctive advantages. As also summarised in Table~\ref{table:related_methods}, classical methods typically assume clean or labelled training data, can only be applied to numerical features, and provide limited interpretability. In contrast, our approach is explicitly designed to operate under data contamination, integrates both numerical and textual features through Doc2Vec, and leverages residual-based scores to deliver ante-hoc explanations. These aspects underline the novelty and uniqueness of our framework in enterprise UEBA scenarios.

While the proposed framework demonstrated strong performance, several limitations should be acknowledged.  First, the amount of labelled anomalies was extremely limited, which prevented the use of standard supervised metrics and restricted baseline comparisons. Second, the reliance on Doc2Vec embeddings for process features, while efficient, may obscure rare but highly informative tokens.  Finally, the evaluation was performed on sensitive institution-specific data that cannot be released, which constrains reproducibility.

These limitations point to several directions for future research. A deeper and systematic comparison with standard unsupervised baselines (e.g., Isolation Forest, LOF, one-class SVM) remains a priority,   ideally using a public benchmark dataset such as the CERT Insider Threat dataset \cite{glasser2013bridging} or datasets with more textual features. Future work should also explore more expressive text encoders, such as transformer-based or recurrent architectures, especially for longer and more natural textual features (e.g., email content, web addresses). In parallel, advancing explainability remains essential. Our residual-based approach already highlights which features are most difficult to reconstruct, offering ante-hoc insights into the features that most influence anomaly detection. Post-hoc techniques such as Shapley values \cite{lundberg2017shap} or LIME \cite{ribeiro2016lime} could complement this by accounting more explicitly for feature interactions and correlations, providing analysts with a second layer of explanation to validate and enrich our residual-based insights. Moreover, counterfactual analyses \cite{wachter2018counterfactual} further extend this perspective by showing what minimal behavioural changes would render an anomalous user or entity normal, which could help analysts perform root-cause analysis and guide remediation strategies. Finally, ensemble strategies that combine complementary anomaly detectors offer a promising route to enhance robustness and reliability in practical deployments.

\section*{Data Availability}
The real world enterprise logs analysed in this study were provided by a financial institution under confidentiality and security agreements and are not publicly available. Aggregate statistics and the synthetic anomaly generation procedure are described in Section~\ref{sec:methods} and Subsection \ref{sec:results-sim}.

\section*{Author contributions}
Conceptualization, J.F, I.O-F, N.M.V and M.S; software, J.F; validation, I.O-F, N.M.V and M.S; formal analysis, J.F, I.O-F, N.M.V and M.S; investigation, J.F; resources, I.O-F, M.S; data curation, J.F; writing—original draft, J.F, I.O-F, N.M.V; writing—review \& editing, I.O-F, M.S, and N.M.V; visualization, J.F; supervision, I.O-F, M.S, and N.M.V; project administration, I.O-F; funding acquisition, I.O-F, M.S.

\section*{Acknowledgments}
This work was partially supported by the European Union's Horizon Europe Research and Innovation programme under the project PRESERVE (Grant Agreement 101168309); the Ayudas Cervera para Centros Tecnol\'ogicos grant of the Spanish Centre for the Development of Industrial Technology (CDTI) under project CICERO (CER-20231019) and by the grant PID2020-118101GB-I00 from the Ministerio de Ciencia e Innovaci\'on (MCIN/AEI/10.13039/501100011033).

\section*{Conflict of Interest}
The authors declare that they have no known competing financial interests or personal relationships that could have appeared to influence the work reported in this paper. The funding sources listed in the Acknowledgments had no role in the study design; in the collection, analysis, or interpretation of data; in the writing of the report; or in the decision to submit the article for publication.

\bibliographystyle{reference_style}
\bibliography{bibliography_final}
\end{document}